# Liquid Collection on Welwitschia-Inspired Wavy Surfaces

Yuehan Yao[1], Christian Machado[2], Youhua Jiang[2], Emma Feldman[3],
Joanna Aizenberg[4, 5*], Kyoo-Chul Park[2*]

1. Department of Materials Science and Engineering, Northwestern University, Evanston, IL 60208, USA;
2. Department of Mechanical Engineering, Northwestern University, Evanston, IL 60208, USA;
3. Department of Chemical and Biological Engineering, Northwestern University, Evanston, IL 60208, USA.
4. John A. Paulson School of Engineering and Applied Sciences, Harvard University, Cambridge, MA 02138, USA;
5. Wyss Institute for Biologically Inspired Engineering, Harvard University, Cambridge, MA 02138, USA;

**Abstract:** Hydrophobic (HPo) surfaces for atmospheric water harvesting applications require sophisticated wettability and microstructure patterns, which suffer from high cost and low durability against severe mechanical and environmental degradations. Inspired by the leaves of *Welwitschia mirabilis*, a long lifespan desert-living plant, we present a robust superhydrophilic (SHPi) surface design which is capable of enhancing both dew condensation and fog capture. The surface consists of a parallel-aligned millimetric wavy topography on top of which random nanostructures are grown. By studying the liquid dynamics, we have shown a unique mechanism of water transport on the SHPi wavy surface by droplet splitting and Laplace pressure gradient. It has been revealed that the efficient water transport on the SHPi wavy surface synergistically interacts with the enhanced diffusion flux of water vapor on the convex regions and significantly improves the efficiency of filmwise condensation. We have further demonstrated that wavy features enhance the probability of fog droplet collisions by disturbing the air flow and increasing the mean free path of fog droplets.

**Introduction**

Dew condensation and fog capture are the two main mechanisms of atmospheric water harvesting.[1-3] As a phase change heat transfer phenomena by nature, dew condensation on a supercooled surface consists of nucleation, droplet growth, and condensate removal.[4, 5] Although nucleation is energetically preferred, superhydrophilic (SHPi) surfaces, on which condensation occurs in a filmwise manner, are believed to be inferior to hydrophobic (HPo) and superhydrophobic (SHPo) surfaces due to the thermal insulating effect of the condensate film.[6-9] However, dropwise condensation can fail on SHPo surfaces under a large subcooling and a high condensation rate, because droplets inevitably nucleate within the nanoscale surface textures, which impales the Cassie-Baxter wetting state and causes the surface flooding problem.[10, 11] Condensate on the flooded surfaces is subject to significant retention force due to the surface roughness, and the condensation efficiency can be largely deteriorated because of the immobile droplets.[12, 13]

On the other hand, water harvesting by capturing fog droplets is not only an interfacial process, but also a problem of fluid dynamics. Fog droplets (1-40 μm in diameter) in the air flow collide with the solid surfaces at first, and are then transported and collected.[14, 15] In the first step, the macroscopic geometry of the fog collectors influences the air flow field near the surface and hence the probability of fog droplets colliding with the surface.[16, 17] Microscopically, it has been shown that the surface wettability needs to be carefully chosen, since the SHPi mesh suffers from the clogging problem which reduces the permeability of the collector, whereas the deposited fog droplets on the SHPo mesh can reenter the air flow without being effectively captured.[18, 19] However, for the water transport step, studies on the fog capturing behavior of single wire systems revealed that the onset time needed for collecting fog drops in the reservoir is minimized on the SHPi surfaces because of the spontaneous water spreading.[20, 21] Therefore, a surface design with proper macroscopic geometry and surface wettability to optimize both fluid dynamics and liquid transport is yet to be fully developed.[22, 23]

Here, we report the surface design with millimetric wavy patterns inspired by the leaves of *Welwitschia mirabilis*, a long life-spanned desert-living plant, for both efficient dew condensation and fog capturing. For condensation, we employ the diffusion-enhanced droplet growth mechanism on the convex surface features discovered in our previous studies.[24, 25] We further study the droplet transport mechanism on the convex surfaces with various wettabilities. We demonstrate that, different than on the extensively-studied flat surfaces, the SHPi convex surfaces benefit from a unique droplet splitting mechanism which keeps the thickness of condensate to be effectively zero. The water mobility is further enhanced by introducing the curvature gradient to the surface geometry so the condensate is spontaneously transported away from the convex regions driven by the Laplace pressure. For fog capturing, we show that the wavy features disturb the air flow so the probability of droplet collision with the surface increases. The SHPi wettability effectively retains the captured fog droplets, and transports them to the reservoir.

**Methods**

**Macroscopic Surface Patterning.** Surfaces with millimetric feature patterns were fabricated by pressing a thin aluminum sheet (Al 1100, 0.127 mm in thickness, McMaster-Carr) between a pair of 3D-printed molds (Form 2 3D printer and clear resin, Formlabs). Two types of surface geometry, wedge-shape and wavy (See Results and Discussions for more details), were used.

**Surface Modifications.** The patterned surfaces were then cleaned by oxygen plasma (PC 2000 Plasma Cleaner, South Bay Technology) for 1 min to remove the organic contaminants. By introducing hydrophobic coatings and/or nano-textures, the surface wettability can be changed to superhydrophilic (SHPi), hydrophobic (HPo), and superhydrophobic (SHPo). (1) The hydrophobic samples were made by immersing the cleaned aluminum sheets in 1 wt% solution of fluoroaliphatic phosphate ester fluorosurfactant (FS-100, Pilot Chemical) in ethanol at 70°C for 30 minutes. After the liquid-phase hydrophobic coating treatment, the samples were then rinsed with DI water to remove the excessive ethanol on the surfaces, and were further dried with compressed air. (2) The superhydrophilic samples were made by boiling the cleaned surfaces in water at 100°C for 30 min. Aluminum undergoes the Boehmite process in boiling water, and nano-structures grow on the surfaces. (3) The superhydrophobic samples were made by the Boehmite process as the first step, and then applying the hydrophobic coating. The boiled samples were cleaned by ethanol (200 proof, Fischer Scientific) and dried by compressed air to remove the excessive water before the hydrophobic treatment. The static contact angles of water on the SHPi, HPo, and SHPo surfaces were measured to be ~0°, 114°, and 152°, respectively.

**Liquid Deposition.** The droplet instability was studied by depositing the liquid drops on the apex of the wedge-shaped hydrophobic surfaces with two apex angles (11° and 100°, CA = 114°). By making a 1:3 mixture of 2-propanol and water (volume ratio), the CA can be modified to be 49.6°. A syringe pump (Pump 11 Elite, Harvard Apparatus), a syringe (3 mL, Becton Dickinson), and a stainless steel needle (0.41 mm OD, McMaster-Carr) were used to dispense liquid drops (DI water or IPA/water mixture) at an infusion rate of 1 μL/sec. To minimize the kinetic energy of the drops, the wedge surface was horizontally placed close to the needle so the distance that the drops fell before touching the surface was small (about 1 mm). A high-speed camera (Photron Mini AX) with a macro lens (Nikon AF-S NIKKOR 85mm f/1.8G) was used to record the cross-sectional view of the deposited droplets at a frame rate of 5000 fps.

**Visualization of the Water Meniscus.** The profile of the water meniscus on the superhydrophilic wavy surface was visualized under a reversed confocal laser scanning microscopy (Leica SP5, 5× objective lens). Rhodamine 6G (99%, Sigma-Aldrich), a typical dye for fluorescence probes, was dissolved in DI water at a concentration of 1 mg/L. The solution was then dispensed on the horizontally-placed superhydrophilic wavy surface to ensure the sample is uniformly wetted. The sample was then gradually tilted until the waves are vertically oriented, and the excessive liquid was removed from the surface by gravity. The sample was placed upside-down on the stage, and was scanned by a 520 nm laser beam which has a shorter wavelength than the absorption maximum of Rhodamine 6G at 530 nm, and fluorescence emission within 550-590 nm was collected to form the image. To obtain the cross-sectional profile of the meniscus, the scanning was conducted under the $z$-stack mode with a total $z$-volume of 2.5 cm and a step size of 1 μm to cover the entire range from the valley to the peak of the sample.

**Condensation.** The condensation experiments on the wedge-shape and the wavy surfaces were done in a customized environmental enclosure. The interior temperature ($T_{air}$) was 22°C. To control the water content in the gas phase, an ultrasonic humidifier (IVADM45, Ivation) was placed inside the chamber and was connected to an external PID controller (Model 5200, Electro-Tech Systems) so the relative humidity ($RH$) was kept at 80%. The samples were attached to a cascaded cooling stage by double-sided copper tapes (3M™ Scotch Double Sided Conducted Copper Tape, 12.7 mm wide and 0.04 mm thick). The cooling stage consists of a liquid cold plate (Aluminum, 4-pass tubes, 305 mm × 197 mm × 16.7 mm, Length × Width ×Thickness, Wakefield-Vette) as the base, and a thermoelectric cooling top plate (CCP Cascades, TECA) which is attached onto the base plate by a thermal pad (0.229 mm in thickness, McMaster-Carr) and thermal paste (Series 7, Protronix). The base plate was connected to an external circulating chiller (7L AP, VMR) at 10°C, and the voltage applied to the top plate was regulated by a digital DC power supply (DCP 305D, Yescom). To avoid condensation before the ambient conditions are stabilized, the surface temperature of the samples ($T_{surf}$) was kept at 25.0°C by applying a negative bias to the top plate. The bias applied to the top plate was then switched to positive, at which $t = 0$ is defined, and the surface temperature reached 2.5°C within 5 min. The condensation processes were recorded by a DSLR camera (Nikon D5500) and a macro lens (Nikon AF-S NIKKOR 85 mm f/1.8G). A schematic of the condensation setup is shown in Fig. 1A.

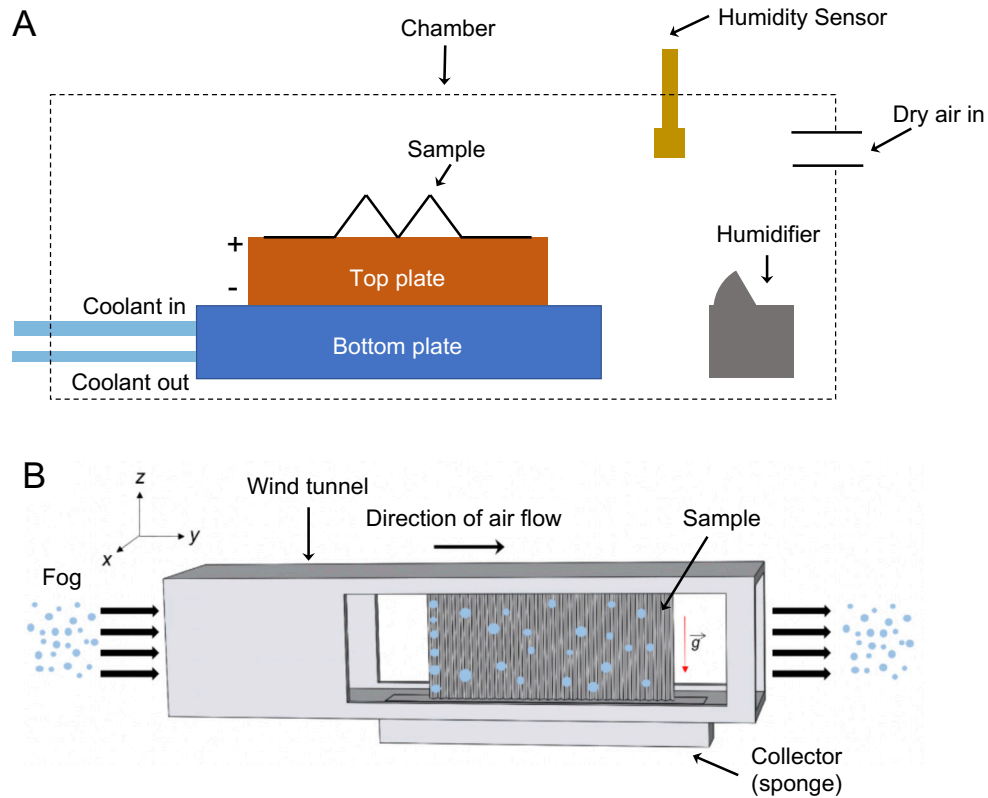

**Figure 1.** Schematic of the experimental setups for (A) dew condensation, and (B) fog capture.

**Fog Collection.** Fog collection experiments were performed in a custom-built, 3-D printed wind tunnel (Formlabs Form 2 clear resin). The wind tunnel was housed inside an enclosed chamber that maintained high humidity (~95%) via an ultrasonic humidifier (IVADM45, Ivation). At the outlet of the wind tunnel, a fan (AC Infinity) was placed to create airflow within the wind tunnel. The airspeed could be controlled by varying the voltage to the fan, via using a variable voltage transformer (VARIAC). At the inlet of the wind tunnel, a flow stabilizer was attached to ensure uniform flow upon entering the wind tunnel. Immediately after the flow stabilizer, an additional ultrasonic humidifier was connected to the interior of the wind tunnel through a custom-made attachment. Midway along the wind tunnel, the samples were suspended by a 3-D printed cap with the same wavy geometry. The wavy samples were maintained to have the same projected dimensions, 6 cm x 10.5 cm. Underneath the suspended samples was a foam piece used to absorb water shed by the wavy samples during fog collection. Experiments were performed at three windspeeds, 3 m/s, 2 m/s, and 1 m/s, and included all four wettabilities, SHPi, HPi, HPo, and SHPo. Furthermore, a HPi flat sample of the same projected area was also tested and compared to the wavy samples. Before experiments were run, a control was performed to quantify the mass of water absorbed by the foam piece without any sample. This value was used to subtract from the mass gain of water by the various samples. Each trial was performed three times to ensure repeatability of the data. A schematic of the condensation setup is shown in Fig. 1B.

**Numerical Simulation of the Fog Capture on the Wavy Surfaces.** The fog capture on the wavy surfaces is numerically simulated by a two-step method using COMSOL Multiphysics. A 2D model with the wavy geometry (72 mm long) positioned at the center of a rectangular wind tunnel (60 mm × 200 mm, width × length) is built. The background velocity field at steady state is simulated at first by using the laminar flow model. The air velocity at the inlet is kept to be 1 m/s, and the pressure at the outlet is kept zero. No-slip boundary condition is assumed on the side walls of the tunnel and on the wavy surface. A time-dependent particle trajectory simulation is then run by releasing water droplets (10 microns in diameter, 40 particles per 1μs) with an initial velocity of 1m/s at the inlet. The fog droplets are assumed to be frozen once they touch the wavy surface.

## Results and Discussions

### Droplet Stability on a Wedge-Shape Convex Surface

Due to the constraint of minimum surface energy, the equilibrium shape of a droplet resting on a flat surface is a spherical cap when the capillary force dominates.[26] As shown in Fig. 2A, the maximum thickness of the liquid cap ($h_{max}$), and the water-solid interface area ($A$) is given by:

$$h_{max} = \left(\frac{3}{\pi}\frac{1-\cos\theta}{2+\cos\theta}\right)^{\frac{1}{3}} V^{\frac{1}{3}} \quad (1)$$

$$A = \pi a^2 = \pi \left(\frac{3}{\pi}\frac{1-\cos\theta}{2+\cos\theta}\right)^{\frac{2}{3}} \left(\frac{\sin\theta}{1-\cos\theta}\right)^2 V^{\frac{2}{3}} \quad (2)$$

where $V$ is the volume of the droplet, $\theta$ the apparent contact angle, $a$ the radius of the contact area. Fig. 2B shows that $h_{max}$ increases as $\theta$ increases, while $A$ decreases as $\theta$ increases. Greater $h_{max}$ implies a thicker layer of liquid and hence greater thermal insulation. Fig. 2C shows the droplet on a supercooled hydrophobic bump after 90 min of condensation ($T_{surf}$ = 8.5°C, $T_{air}$ = 21.2°C, $RH$ = 70%). Droplets at the apex grow too big in size and completely cover the bump. As pointed out in our previous studies, the bumpy geometry enhances condensation by increasing the diffusion of water vapor near the surface feature.[24, 25] However, such enhancement can be compromised by the poor heat transfer of the condensed liquid if it is not transported away efficiently.

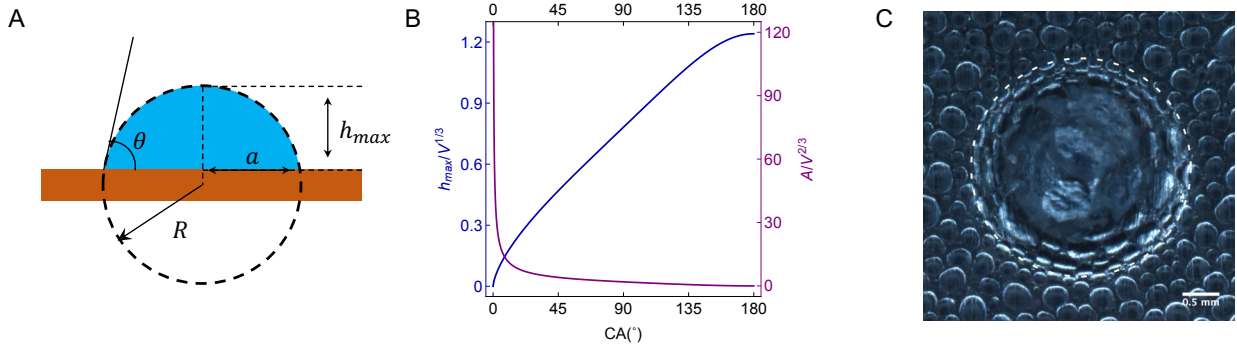

**Figure 2.** (A) Schematic of a spherical cap sessile droplet on the surface. (B) The maximum thickness of the droplet cap ($h_{max}$) increases with the contact angle (CA), and the total water-solid contact area ($A$) decreases with CA. (C) Condensed droplets are pinned at the apex of the hydrophobic bump, making the apex completely covered by water.

To address the condensate coverage issue without losing the diffusion mechanism, we try to understand the dynamics of a droplet resting on a simple convex surface which consists of two half-planes tilted at an angle ($\alpha$). The geometry can be changed by varying the $\alpha$ value (11°, 100°, and 180°). Fig. 3 shows the dynamics of a impacting droplet and its interaction with the hydrophobic wedge. In condensation, droplet statically grows with zero kinetic energy (if the excessive surface energy released by coalescence is ignored). For this reason, the droplets were released at a low height from the wedge apex to minimize their velocity. Three modes of interactions can be observed depending on the combination of the apex angle ($\alpha$) and the wettability (CA). On a flat surface, where $\alpha$ = 180°, droplets are able to statically rest on the surface regardless of the magnitudes of CA, giving rise to a spherical cap shape with substantial $h_{max}$ (Non-split, stable mode, boxed in blue). This is also true for both CA = 49.6° and 114°when $\alpha$ = 100°. However, if $\alpha$ decreases to 10°, two different fates of the impacting droplet can be observed. With a low CA (49.6°, HPi surface equivalent), the spreading tendency drags the droplet onto the two half planes, and the droplet spontaneously splits into two halves (Split mode, boxed in red). When CA is high (114°), however, the impacting drop could not stably rest on the apex of the wedge. The contact area when the droplet was the most compressed ($t = 9.8$ μs, $\alpha = 11°$, CA = 114°) is much less compared to when it impacts onto a wedge apex with a larger $\alpha$ ($t = 4.8$ μs, $\alpha = 100°$, CA = 114°). The droplet is thereby much less well-supported by the sharper apex. The mass center of the droplet can easily deviate away from the axis of symmetry as the shock waves propagate in the droplet. The droplet easily falls onto one of the two half planes when the deviation is significant enough, resulting in an asymmetric wetting profile (Unstable mode, boxed in brown). Since the enhancement of water vapor diffusion is the most pronounced at the apex of the convex features, the maximum thickness of the droplet coverage at the apex is of significant implications to the overall condensation efficiency. Fig. 3 marks the $h_{max}$ for various CA and $\alpha$ combinations after the kinetic energy of the droplets has fully dissipated ($t = 75$ μs). It is clear that, for the same CA, decreasing $\alpha$ effectively results in a smaller $h_{max}$. When $\alpha$ is large for droplet to statically sit on the apex, decreasing CA leads to smaller $h_{max}$, while when $\alpha$ is small, increasing and decreasing CA can both leads to $h_{max}$ close to zero, by the droplet splitting and repelling mechanism, respectively.

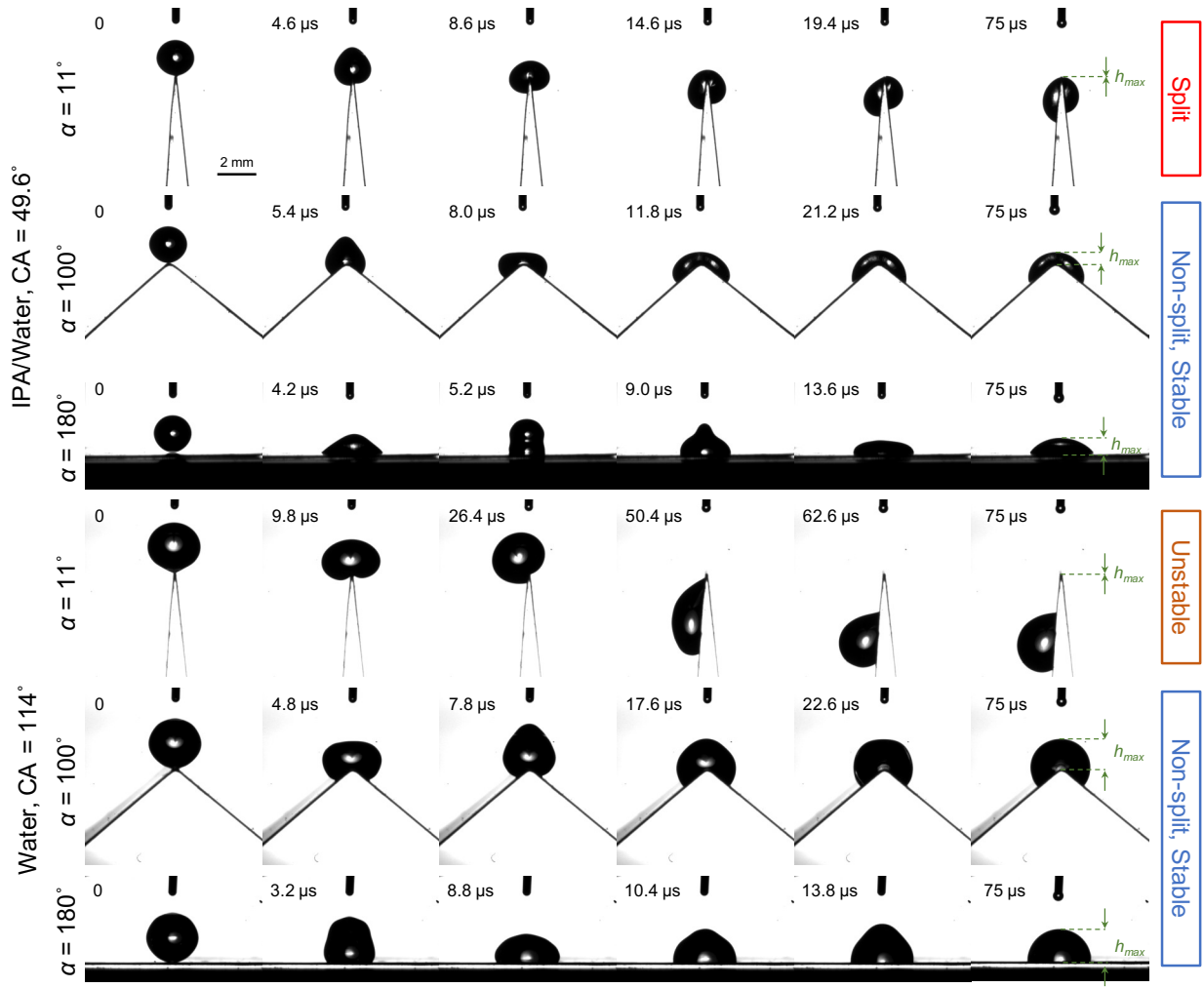

**Figure 3.** Time-lapse images showing the different fates of impacting droplets on the apex of the wedge-shape convex surfaces.

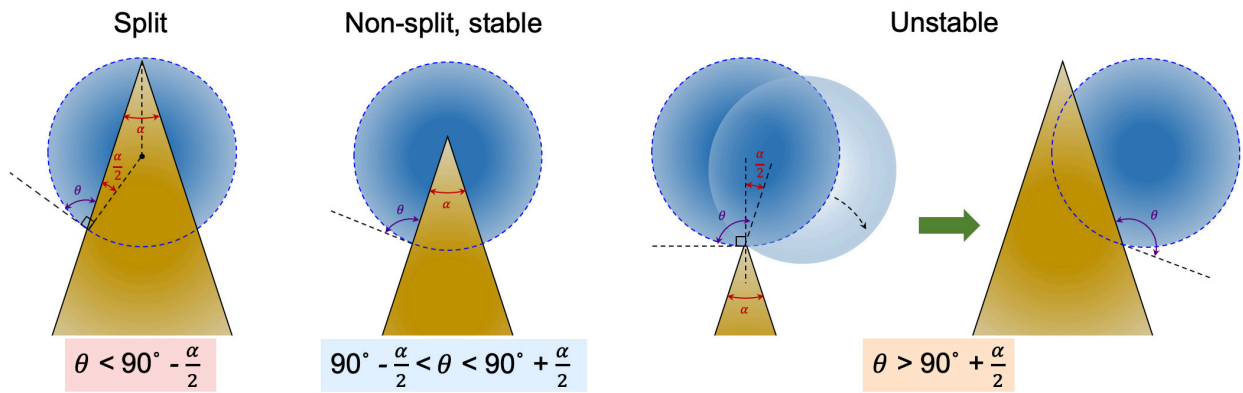

**Figure 4.** 2D schematics of the states of a droplet resting on an apex.

To quantitatively understand the correlation between the fate of the droplet and CA/$\alpha$ combinations, we simplify the problem to 2-dimensional, as shown in Fig. 4. The equilibrium profile of a sessile microdroplet in 2D is part of a circle, since such geometry minimizes the circumference with a constant enclosed area. Two critical relations can be found by this analysis. When CA $< 90° - \frac{\alpha}{2}$, i.e. low CA and low $\alpha$, the surface tension is unable to support the droplet to stably sit on the apex, and splitting will occur. When CA $> 90° + \frac{\alpha}{2}$, i.e. high CA and low $\alpha$, the apex is not wetted by the droplet, and the droplet can fall on the half-plane if the mass center deviates from the axis of symmetry. When CA is intermediate, or $\alpha$ is large enough, droplet can rest stably at the apex, resulting in a large $h_{max}$. This correlation is further verified by numerical simulation of the droplet dynamics. Fig. 5 shows the phase map generated by simulating a wide range of CA/$\alpha$ combination, where the two boundaries between the three different fates of resting droplets are drawn and agree well with the geometric analysis.

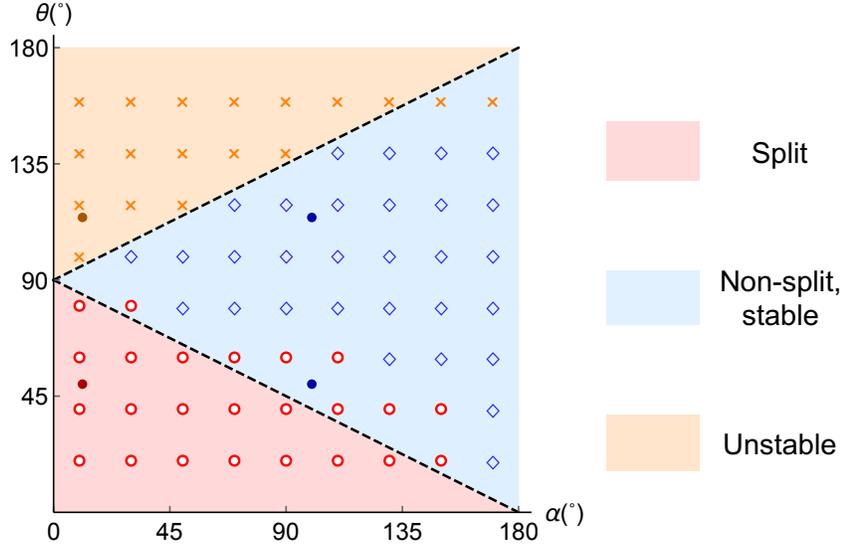

**Figure 5.** Phase map generated by numerical simulations shows the split (red, open circle), non-split stable (blue, open diamond), and unstable (orange, cross) states of the droplet. The solid dots correspond to the experimental results.

**Condensation on the Wedge-Shaped Convex Surfaces**

The droplet instability on the convex surface is verified in condensation. As shown in Fig. 6, condensation happens in distinct manners depending on the surface wettability (CA) and the surface geometry ($\alpha$). On a flat SHPi surface, condensation is uniform without showing spatial preferences. The liquid coverage and the thickness of the thermal-insulating condensate increase as the droplets grow and coalesce, and the condensation rate decreases as time proceeds. On the wedge surface ($\alpha = 100°$) with the same CA, however, droplets primarily nucleate and grow at the apex at first ($t = 20$ min). The wedge apex is not perfectly sharp. Instead, it has a small radius of curvature because of the patterning method. Droplets grow to a critical size in the similar magnitude as this radius of curvature, coalescing with droplets on the two half planes and splitting. In this way, the condensate is effectively removed away, and the apex region is constantly refreshed for condensation cycles.

With the same surface geometry ($\alpha = 100°$), the traditional SHPo surface shows a different behavior of condensation. Similar to the SHPi wedge surface, water preferentially nucleates and grows on the apex ($t = 20$ min). However, the droplets are pinned at the apex and could not be easily transported away. The droplets covering the wedge apex are much larger in size compared to those on the SHPi wedge surface ($t = 40$ min and 60 min). The droplet immobility suggests that the nanoscale roughness of the SHPi surface is impaled and flooding occurs. Water randomly nucleates and coalesces under a high supersaturation ($S \approx 2.9$), and the Cassie-Baxter wetting state transitions to the Wenzel state. Because of the significant pinning, condensed droplets are unable to fall on the half-planes as is observed in the droplet impact experiments.

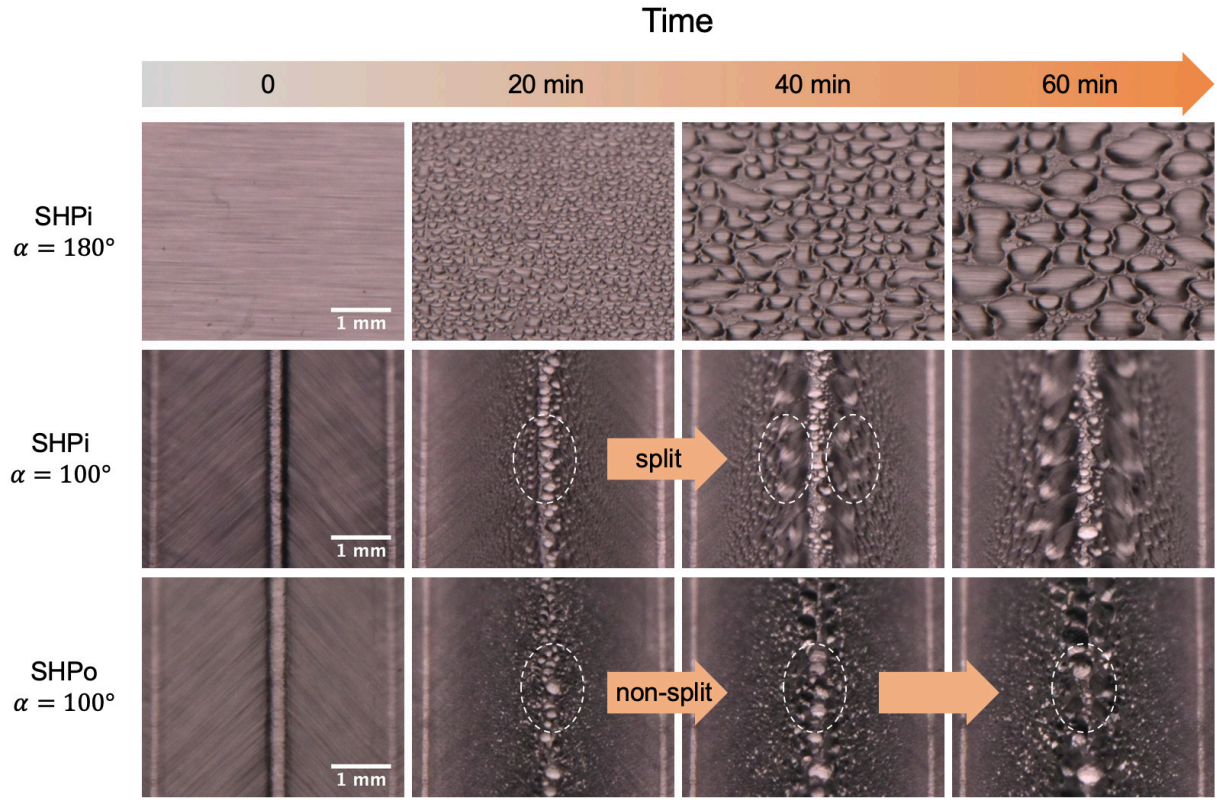

**Figure 6.** Condensation on the flat and the wedge-shape convex surfaces. SHPi surface shows filmwise condensation with appreciable liquid film thickness. The condensate splits on the SHPi wedge surface, making the apex continuously refreshed. Condensed droplets are pinned on the apex of the SHPo wedge surface because of Wenzel wetting.

**Liquid Transport by Continuous Curvature Gradient**

An obvious shortcoming for the SHPi wedge surface for condensation is that, due to the infinite curvature at the apex, the droplets are pinned on the two half-planes where the curvature is zero as shown by the dashed circles in Fig. 6. The splitting mechanism for the wedge geometry does not apply to transport the condensate further away from the apex so the condensate can be collected. To address this issue, we introduce an additional driving force for a continuous liquid transport-collection cycle.

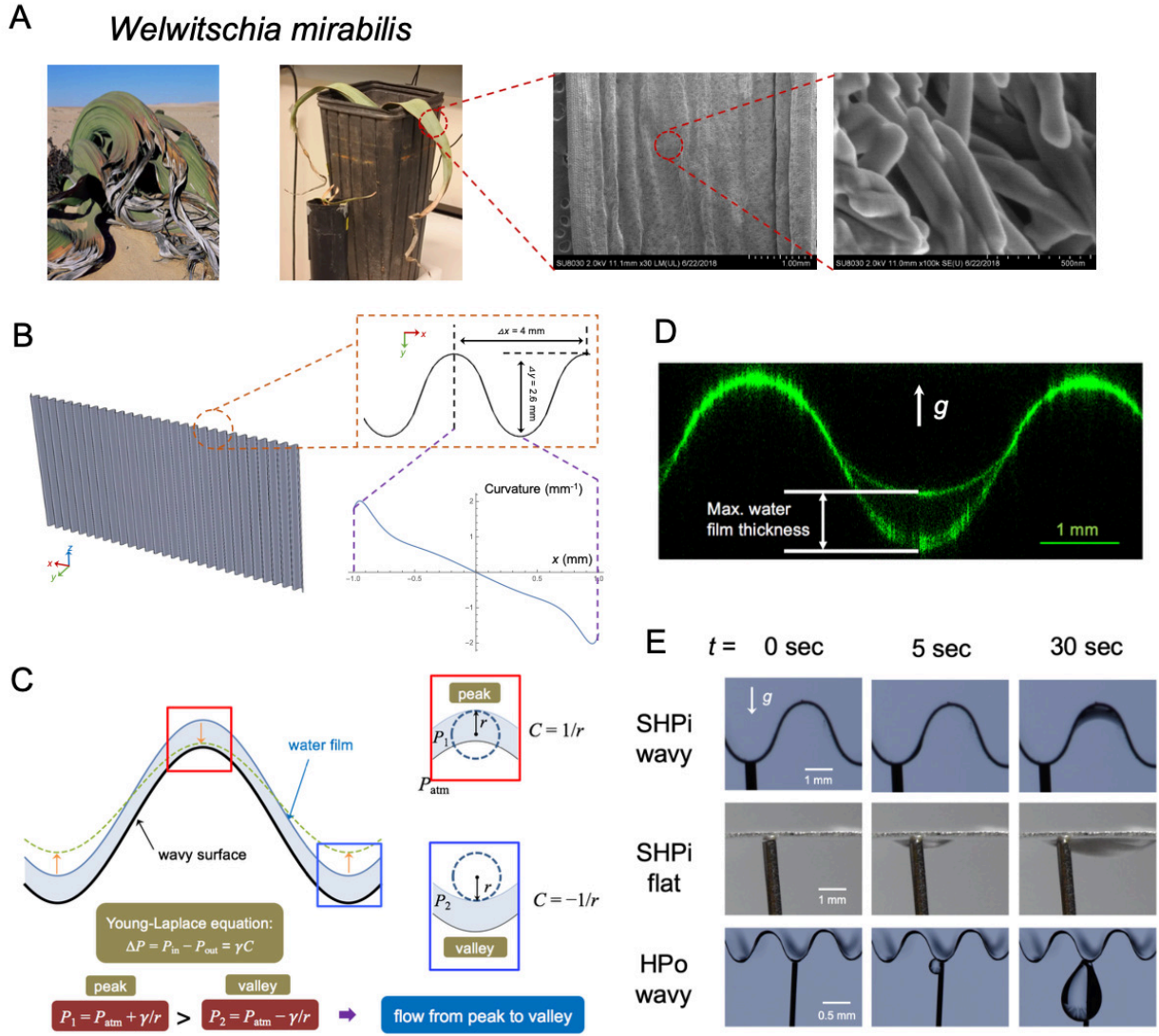

**Figure 7.** (A) The hierarchical surface textures on the leaves of *Welwitschia mirabilis*. (B) Surface design with continuous gradient of surface curvature mimicking the wavy pattern of *Welwitschia*. (C) Schematic of the spontaneous water transport mechanism on the SHPi wavy surface driven by the Laplace pressure gradient. (D) Equilibrium profile of the water meniscus on the SHPi wavy surface. (E) Water infusion and transport on the SHPi wavy, SHPi flat, and HPo wavy surfaces.

*Welwitshcia mirabilis*, a desert-living plant which has a long lifespan of over 1000 years, shows unusual hierarchical structure of its leaves.[27] As shown in Fig. 7A, the leaves show a continuous wavy pattern on the scale of ~1 mm instead of a flat geometry as the first tier, on which the second-tier nanoscale roughness is present. Inspired by this wavy geometry, we modify the wedge shape apex into a wavy pattern with continuous surface curvature. By using the SHPi wettability, the liquid spreads on the surface and well conforms the geometry of the solid substrate. We hypothesize that spontaneous liquid transport will occur because of the Laplace pressure gradient caused by the curvature gradient. We design our surface geometry as a periodic pattern, where the curvature linearly decreases from positive (at the peaks) to negative (at the valleys), as illustrated in Fig. 7B.

Fig. 7C shows the equilibrium profile of water on the SHPi wavy surface. Even though the entire surface is wetted by water, the water then preferentially resides in the valley where the curvature is negative, while the peak region shows a zero thickness of water film. Note that the direction of gravity points upwards, meaning the capillary force outweighs the gravitational force in such length scale. The transport mechanism was also studied by liquid infusion experiments. As shown in Fig. 7D, a needle was in contact with the peaks. By expelling water at a rate of 0.5 μL/s, the liquid spontaneously transports into the valley on the SHPi surface, and the thickness of water at the peak regions where the liquid was infused is negligible. In comparison, on a SHPi flat surface, the liquid accumulates in size, resulting in a substantial thickness which is similar to what is observed in filmwise condensation on a flat surface. On a superhydrophobic surface, however, the droplet grows in size without any lateral motion relative to the peak. The peak was covered by the drop, similar to the pinned droplets shown in Fig. 6.

**Condensation on the Wavy Surface**

The synergy between the diffusion-based droplet growth and the curvature-driven liquid transport was studied by the condensation experiments. To better visualize the motion of condensed liquid, the peak of the wavy surface was dyed by a particle of Rhodamine 6G. As shown in Fig. 8, the solid dye particle was dissolved in the condensed liquid, and shows the path along which the liquid moves. On the SHPi sample, condensation preferentially happens on the peaks, the same as we observed on the wedge-shaped surfaces. Droplets grow and the transport initiates when they are large enough ($t$ = 10 min). The droplets stop their motions when they reach the valleys, where the condensate is collected. It is worth noting that the first the liquid thickness at the peaks is not negligible before the droplets depart from the peaks (*t* < 15 min). This is because droplet pinning occurs at the water front, where the non-wetted region becomes wetted due to the droplet motion. However, such pinning is significantly reduced as condensation proceeds. As shown when *t* = 60 min, most of the peak areas are refreshed without significant condensate thickness. Such enhancement of liquid mobility suggests that water infused in the nanoscale roughness of the SHPi surface, and the condensate is effectively 'self-lubricated'.

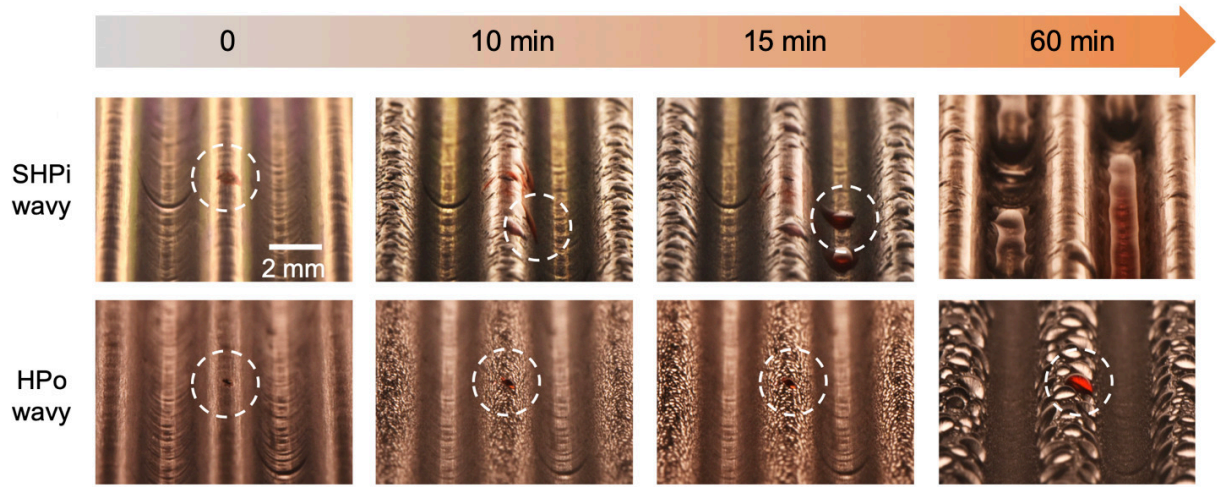

**Figure 8.** Enhanced condensation on the peaks of the SHPi and HPo wavy surfaces. Condensate spontaneously transports and is collected in the valley on the SHPi wavy surface, while the droplets grow and are pinned on the HPo surface.

However, on the hydrophobic and superhydrophobic wavy surfaces, the condensation mode is very different. Because the condensate does not move in a filmwise manner, the liquid collection relies on the motion of each individual droplet. The droplets are therefore pinned on the peaks without moving into the valleys. As the number density and size of the droplets increase, the overall condensation rate slows down, which further inhibits the droplet growth. The Laplace pressure gradient does not apply to such droplets which show very different profile shape than the underlying substrate. The condensed droplets were then detached and rolled down along the peaks because of gravity. As the number of condensation cycles increase, the dropwise condensation mode gradually fails and because of the Cassie-Baxter to Wenzel transition. Different than the high droplet mobility on the SHPi surfaces, the condensate transport did not benefit from the filmwise mode, because the hydrophobic nature of the underlying substrate significantly pins the condensate at the contact line by exhibiting a high energy barrier to be wetted.

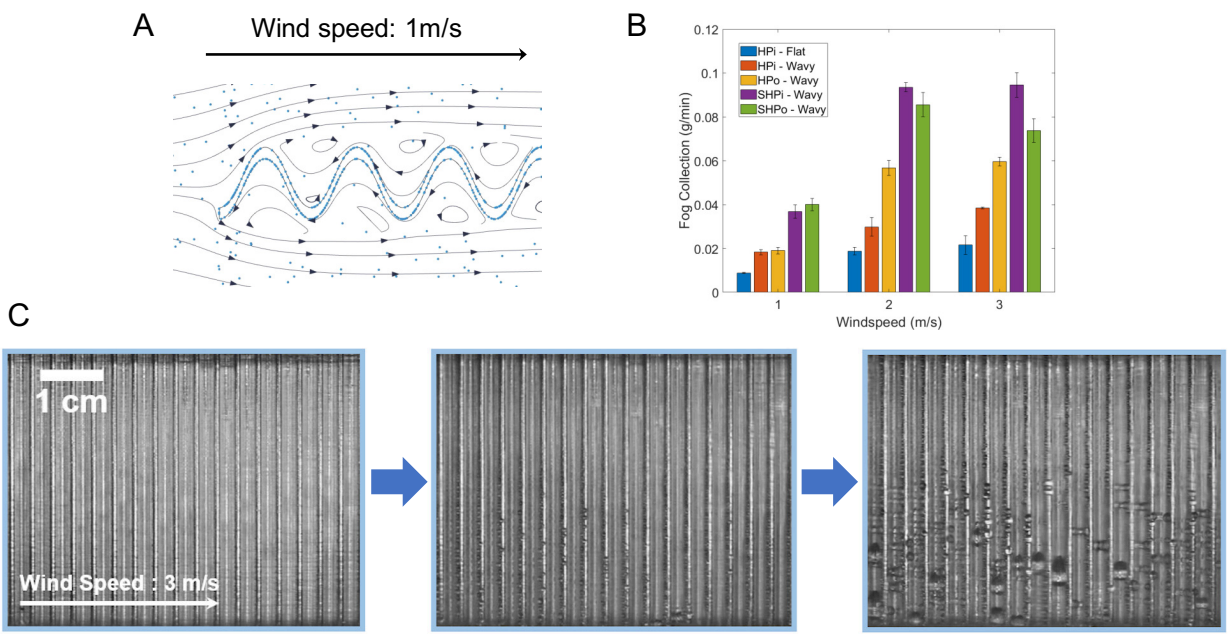

**Figure 9.** (A) A fluid dynamics simulation using COMSOL Multiphysics on a wavy surface at 1 m/s. As the uniform flow initially impacts the surface, separation occurs, resulting in low-velocity areas of circulation that develop within the valleys of the wave geometry. (B) Quantitative fog collection results comparing wavy surface geometries with flat surfaces of the same projected area. Wavy surfaces of four different wettabilities were tested and shown to outperform fog collection on a HPi flat sheet, with a more pronounced increase in collection rate with increasing windspeed. The SHPi wavy surface performed best at the wind speeds tested (1-3 m/s). (C) Mechanism of fog collection on a rigid, hydrophobic wavy surface. Droplets initially deposit on the peaks of the waves, then transport into the valleys as they grow. Once a droplet reaches a critical volume, the droplet sheds from the surface and is collected.

**Fog Collection on the Wavy Surface**

The quantitative results are shown in Fig. 9. As shown, the fog collection rates for all four wettabilities of the wavy surfaces were higher than that of the flat HPi surface. Furthermore, the SHPi wavy surfaces possessed the highest fog collection rate at 2 and 3 m/s. Simulation results of flow streamlines interacting with the wavy surfaces are shown in Fig. 9A. As shown, the separation occurs almost immediately after contacting the surface, resulting in low velocity vortices located in the channels of the waves, while bulk flow is found farther away. The free path of fog droplets near the wavy surface features is significantly elongated by the vortices, and hence the probability for the droplets to collide with the surface and get captured is enhanced. A qualitative assessment of experimental data in Figs. 9B and C shows that droplets in the bulk flow initially deposit on the peaks of waves, coalesce, and are transported into the valleys using the Laplace pressure gradient. There, the droplets continue to grow in a low velocity regime, thus reducing evaporative losses as well as losses due to re-entrainment. The droplets continue to grow until a critical volume is achieved where gravitational forces exceed the retention force dictated by the surface chemistry and morphology. At this point, droplets are shed downwards to the collection apparatus.

**Conclusion**

In all, we studied the optimized combination of macroscopic and micro/nanoscale surface texture as well as the surface chemistry for efficient water condensation from the air. We discovered the distinct modes of water droplet when it sits on the apex of a wedge-shape convex surface. The droplet splitting mechanism, especially, gives an opportunity to the development of efficient filmwise condenser which does not rely on the degradable hydrophobic coatings. However, by combining the enhanced diffusion on the convex surface features, it was found that the droplets were pinned on the two half-planes without being transported further away from the apex, and the condensation efficiency could be potentially compromised. Inspired by the hierarchical surface texture of the leaves of *Welwitschia mirabilis*, we rationally designed a surface profile with continuous curvature gradient. We found that liquid can be spontaneously transported from the peaks to the valleys, following the curvature gradient. By combing the wavy geometry and the superhydrophilic surface wettability, we were able to achieve an efficient condensation of water vapor which outperforms the dropwise condensation on the hydrophobic and superhydrophobic surfaces, which were believed to be superior to the filmwise mode. We envision this study provides physical insights on the phase change and transport phenomenon for a wide range of applications including condensation heat exchangers, desalination, and microfluidics.